\documentclass{aa}
\usepackage{graphicx}

\begin{document}

%

\title{CCD drift-scan imaging lunar occultations: a feasible approach for
sub-meter class telescopes}
\titlerunning{CCD drift-scan imaging lunar occultations}

\author{O. Fors\inst{1}\fnmsep\inst{2}
\and J. N\'u\~nez\inst{1}\fnmsep\inst{2}
\and A. Richichi\inst{3}}

\offprints{O. Fors, \email{ofors@am.ub.es}}

\institute{Departament d'Astronomia i Meteorologia, Universitat de  Barcelona,
Av. Diagonal 647, 08028 Barcelona, Spain
\and
Observatori Fabra, Cam\'{\i} de l'Observatori s/n, 08035 Barcelona, Spain
\and
European Southern Observatory, Karl-Schwarzschild-Strasse 2, 85748 Garching bei
M\"unchen, Germany.\\
}

\date{Received / Accepted}

\abstract{A CCD drift-scanning technique for lunar occultation (LO) observations is
presented. While this approach has been addressed before by Sturmann
(\cite{laszlo94}) for the case of large telescopes, the technical validity has
never been discussed for sub-meter class telescopes. In contrast to Sturmann's
scheme, the proposed technique places the CCD in the image plane of the
telescope. This does not represent a problem in the case of small telescopes, where
the practical angular resolution attainable by LO is not limited by aperture
smoothing.  Photon-generated charge is read out at millisecond rate on a
column-per-column basis, as the diffraction pattern of the occulted star is being
tracked. Two LO events (\object{SAO 79031} and \object{SAO 77911}) were observed to
demonstrate the feasibility of the method. Data  analysis was carried out and
no angular diameter the observed stars were resolved. We show, however, that the
technique could be useful for close binary detections with small telescopes. A
discussion of the limiting resolution and magnitude imposed by our instrumentation
is carried out, showing that drift-scanning technique could be extended to 1-2m
telescopes for stellar diameter determination purposes. Finally, we point out that
the technical demands required by this technique can be easily met by most small
professional observatories and advanced amateurs.
\keywords{Instrumentation: detectors --  Techniques: high angular resolution --
Occultations -- Stars: fundamental parameters -- (Stars:) binaries: general}
}

\maketitle

\section{Introduction} \label{introduction}

Along with eclipses, lunar occultations (hereafter LO) are the oldest astronomical 
phenomena ever recorded. Since the first high-speed photoelectric observations
(Whitford \cite{whitford39}) LO have become one of the highest angular resolution
techniques available in visible and infrared astronomy. It provides the opportunity
of obtaining milliarcsecond resolution, far beyond the seeing and the diffraction
limits of even the largest telescopes.  Also note that optical LO is one of
the few astronomical observations which can be successfully performed from
light-polluted areas, since the natural background of the Moon overcomes any other
background source. The purpose of observing such events has changed along
centuries: geographical longitude calculation, Earth rotation studies, investigation
of close binaries and stellar angular diameters. While timing applications have been
largely overcome by atomic clocks and GPS, LO still plays an important role in the
direct establishment of fundamental stellar quantities. Tests of stellar atmosphere
models via the derived effective temperatures (Richichi et~al. \cite{richichi98})
and studies of stellar pulsation and circumstellar shells (Richichi et~al.
\cite{richichi88}) are just a few examples of them.

In the 1970s, several intensive LO programs were started in the visible
(Nather \& Evans \cite{nather70}, Ridgway et~al. \cite{ridgway74}). Later infrared
programs were also initiated (Ridgway \cite{ridgway80}, Richichi \cite{richichi87}).
The usual detectors for all programs have been high speed photometers, with
different photomultiplier technology depending on observing frequency (GaAs for
visible and InSb for NIR). For the  visible, we call such systems photoelectric
photometers (PEP). PEP technical basis is widely documented in the literature
(Henden \& Kaitchuck \cite{henden90}).

CCDs have never been used in a routine way for LO observational programs. During
the last two decades, technical specifications of CCDs have been constantly
improved. Despite this rapid development, most current research grade cameras
are still not able to meet read out speed which LO work demands (1\,ms sampling
time), while operating at low read-out noise (RON) and high digitization
resolution mode. The situation in the near-IR domain is different, with
panoramic arrays being used for LO observations thanks to the possibility of
implementing fast read-out schemes on subarray sections. An example is the
MAGIC camera (Herbst et~al. \cite{herbst93}) at the Calar Alto observatory
(Richichi \cite{richichi96}). Even in this successful case,  the achieved
sampling interval time is several times slower than the one achieved by
photometers.

Recently, with the advent of adaptative optics (AO), a few expensive state-of-the-art frame transfer imagers have been released for being assembled in
wavefront sensor systems (Ragazzoni \cite{ragazzoni98}). These devices can operate
above 3Mpixel/sec rates at moderate RON regime, meeting most of LO requirements.
While they will play a key role in the near future as a medium-cost solution for
fast photometry programs, their cost is currently too high for most professional and
high-end amateur astronomers. 

In this paper we show how low-cost full-frame CCDs can be considered as an
alternative for LO detection, when operated in drift-scanning mode.

In Sect.~\ref{cons} we state the main observational constraints when observing
LO events. In Sect.~\ref{app}  we compare the traditional observing method
based on PEP systems, to the the drift-scanning technique with CCDs that we
propose. In Sect.~\ref{snr_evaluation}, we present signal-to-noise ratio
(hereafter SNR) constraints for LO as an indicator of data quality, and a
theoretical estimation of lightcurve SNR for both detector systems is performed. In
Sect.~\ref{obs} we present in detail two observations by the drift-scanning
technique. Finally, a brief description of the data analysis procedure is shown in
Sect.~\ref{res}. Results from fits with both an unresolved source and a binary
source model are discussed.

\section{Limiting angular resolution} \label{cons}

As for every astronomical observation, a LO lightcurve is distorted by several
non-ideal observing factors, each one having different characteristics.
Typically, we can distinguish between two aspects which limit the data quality:
instrumental effects and random errors. While the former are usually  deterministic
and can be removed {\it a posteriori} if we have an accurate calibration of our
instrumental equipment, the latter is associated with stochastic processes with a (presumably) known noise distribution function.

We can also consider the SNR of the lightcurve as an indicator of achievable
angular resolution. Actually, this is the key parameter when evaluating data
quality, since it establishes how well deterministic distortions can be
restored until their effects vanish into noise fluctuations.

In Sects.~\ref{constraints1} and \ref{constraints2}, we outline each one of the two mentioned limits on data quality.

\subsection{Instrumental effects} \label{constraints1}
When recording stellar interference fringes of an occulted star, the resolution
limit, i.e. the minimum resolvable angle, $\phi_m$, is fixed by several
instrumental parameters. In the nearly point-like source domain, three of them
apply among others: aperture of the telescope $D$, filter bandwidth, $\Delta
\lambda$, and integration time, $\tau$. From Fresnel diffraction pattern and
geometry considerations all three effects can be expressed as (Sturmann
\cite{laszlo97}):
\begin{eqnarray}
   \phi_\mathrm{m}&\cong&0.54(D+v\tau) \label{d}\\
   \phi_\mathrm{m}&\cong&0.158(\Delta \lambda)^{1/2} \label{filter},
\end{eqnarray}
where $\phi_m$, $D$, $\Delta\lambda$ and $\tau$ are expressed in $\mathrm{mas}$,
$\mathrm{m}$, $\mathrm{\AA}$ and $\mathrm{ms}$, respectively. Finally, $v$
represents the speed of the diffraction pattern in $\mathrm{m\,ms^{-1}}$. 

From Eq.~\ref{d} we see that large telescopes and long integration times, in
spite of increasing SNR, blur high frequency information. This is one of the
few cases in which the size of the telescope plays against the observer. On
the other hand, with smaller $D$ and shorter $\tau$ the resolution is
preserved, but the SNR is decreased, making more difficult the {\it a posteriori}
removal of instrumental distortions, and restricting  observations only to
bright stars. For $m_{\mathrm{V}}$$\le$5 stars, this trade-off relation
balances to optimal SNR for about 1\,m telescope and 1\,ms integration time.
For a typical value of $v$ of 0.5$\mathrm{m\,ms^{-1}}$, the above relation
yields $\phi_m$=0.8$\mathrm{mas}$.

Since diffraction is a wavelength-dependent phenomenon, polychromatic
observations introduce an additional distortion in the lightcurve. As seen in
Eq.~\ref{filter}, the magnitude of this effect depends on filter bandwidth.
Again, we find a trade-off between $\Delta\lambda$ and recorded SNR, which must
be balanced properly. Considering, as in our case, a Johnson R filter
($\Delta\lambda\sim$580$\mathrm{\AA}$) we obtain $\phi_m$=3.8$\mathrm{mas}$. 

As stated before, given sufficient SNR it is possible to deconvolve for these
deterministic effects on the lightcurve, and achieve angular resolution much
smaller than the formal limits of Eqs.~\ref{d} and \ref{filter}.

\subsection{Random errors} \label{constraints2}

Depending on the detection technique and the temporal resolution, different
noise sources appear. In the particular case of LO work, we should consider the
following as dominant.

Firstly, in most astronomical situations, the detection process is Poisson
distributed, i.e. the probability of obtaining a realization of intensity $k$
coming from a source of mean flux $\mu$ is given by the Poisson probability:
\begin{equation}
{\bf P}(k|\mu) = e^{-\mu}\;\;\frac{(\mu)^{k}}{k!}
\label{poisson}
\end{equation}
with an uncertainty over every realization $\sigma$=$\sqrt{\mu}$. It is
important to note that Poisson noise, also known as shot noise, is inherent to
light nature, and does not depend on detector used.

Secondly, in some detectors, as CCDs, the realization $k$ is read by the
electronics of the detector, and a Gaussian RON of zero mean and standard
deviation $\sigma_{\mathrm{CCD}}$ is introduced. The probability of obtaining a
particular realization $m$ from $k$ is
\begin{equation}
{\bf P}(m|k) =
\frac{1}{\sqrt{2\pi}\sigma_{\mathrm{CCD}}}\exp\left[-\frac{(m-k)^2}
{2\sigma_{\mathrm{CCD}}^2}\right]
\label{gauss}
\end{equation}
If, as in the case of CCDs, both processes in Eqs.~\ref{poisson} and
\ref{gauss} are at play, the Poisson+Gaussian compound probability
of obtaining a realization $m$ given the mean $\mu$ and all its possible
Poisson realizations $k$ is (N\'u\~nez \& Llacer \cite{jorge93}):
\begin{equation}
{\bf P}(m|\mu) = \sum_{k=0}^{\infty}
\frac{1}{\sqrt{2\pi}\sigma_{\mathrm{CCD}}}\exp\left[-\frac{(m-k)^2}
{2\sigma_{\mathrm{CCD}}^2}\right]
e^{-\mu}\;\;\frac{(\mu)^{k}}{k!}
\label{poisson+gauss}
\end{equation}
Thirdly, the light wavefront is distorted due to inhomogeneities in the index
of refraction $n$. This random fluctuation in $n$ makes recorded intensity vary
temporally and spatially. This is normally referred to as scintillation noise.
The intensity after scintillation can be approximated as a log-normal
distribution:
\begin{equation}
{\bf P}(m|k) =
\frac{1}{\sqrt{2\pi}bm}\exp\left[
-\frac{(\ln\frac{m}{k}-\frac{b^2}{2})^2}{2b^2}\right]
\label{lognormal}
\end{equation}
with $b=\sqrt{ln(\sigma_{\mathrm{sc}}^2+1)}$, where $\sigma_{\mathrm{sc}}^2$ is
the scintillation index, which characterizes the strength of the turbulence.

Likewise the former Poisson+Gaussian case in Eq.~\ref{poisson+gauss}, the
Poisson+Log-normal compound probability of obtaining a realization $m$
given the mean $\mu$ and all its possible Poisson realizations $k$ is:
\begin{equation}
{\bf P}(m|\mu) = \sum_{k=0}^{\infty}
\frac{1}{\sqrt{2\pi}bm}\exp\left[
-\frac{(\ln\frac{m}{k}-\frac{b^2}{2})^2}{2b^2}\right]
e^{-\mu}\;\;\frac{(\mu)^{k}}{k!}~,
\label{poisson+lognormal}
\end{equation}
which, as derived in Sturmann (\cite{laszlo97}), yields an uncertainty
\begin{equation}
\sigma_{\mathrm{m}}^2 = \sigma_{sc}^2\overline{m}^2+\overline{m}
+\sigma_{\mathrm{CCD}}^2
\label{sigma_poisson+lognormal+gauss}
\end{equation}
where $\overline{m}$ is the mean number of photons detected in an
integration time. We have included the Gaussian contribution from CCD RON 
accounted by $\sigma_{\mathrm{CCD}}$ in [$\mathrm{photons}$].

The right-hand terms in Eq.~\ref{sigma_poisson+lognormal+gauss} are
scintillation, Poisson and Gauss noise contributions, respectively. For usual
intensity ranges in LO, lightcurve SNR will be marginally affected by CCD
readout noise $\sigma_{\mathrm{CCD}}$. If, as usual, turbulence is not
negligible ($\sigma_{\mathrm{sc}}\neq$0), the scintillation factor must be
taken into account, becoming dominant in the high intensity regime. 

\section{Possible approaches for recording LO} \label{app}

\subsection{Traditional method - Photoelectric photometers (PEP)} \label{pep}

Most LO work at visible wavelengths has been conducted by photoelectric
instruments. These devices are usually based on a single thermoelectrically
cooled GaAs cell detectors. They are placed at the telescope focal plane. The
size of their field of view must be large enough to collect all photons from
the seeing-affected  stellar image, but at the same time small enough to
include as less sky background as possible. One of the more remarkable
features of PEP systems is their effective lack of read out noise: they are
nearly pure photon counting devices. 

\subsection{Proposed method - CCD operating in drift scan mode} \label{tdi}
The conventional use of a CCD device is the operation in stare mode in which
the CCD chip is read out at the end of the exposure. Once the shutter is
closed, the charge generated by the incident light on the surface of the CCD is
converted to digital numbers, in column per column basis, as the clocked charge
moves through a serial register. This has been the usual operating mode in
astronomy for years.

However, other modes can be considered since the clocking rate $\Delta$ can
normally be specified by user. Typically, one has three options:
$\Delta=\Delta_0$, $\Delta<\Delta_0$ or $\Delta>\Delta_0$, where $\Delta_0$ is
the sidereal rate. 

In the first case, the acquired data appear as point-like sources provided
clocking charge direction coincides with star motion over chip and telescope
tracking system is disconnected. In the second case, in order to record
point-like images it is necessary to have the camera properly aligned and to
slow down telescope tracking. These two variants in scanning mode are usually
referred as drift-scanning and time delay integration (TDI). This is the way
several meridian circles (e.g. Stone et~al. \cite{stone96}) and Schmidt
cameras (e.g. Sabbey et~al. \cite{sabbey98}) observe for fast sky coverage at
moderate limiting magnitude. 

The third case is the one we adopt for observing LOs. $\Delta$ can be chosen
according to rate and magnitude of the event to be recorded. The detector does
not need to have a specific orientation thanks to the ability to track with the
telescope. Thus, the stellar image remains stationary over the chip while
photogenerated charge is clocked through the serial register at the desired rate
$\Delta$. It is interesting to note that our technique is, in a sense, based on
the same principle originally proposed by MacMahon (\cite{macmahon08}) to observe
lunar occultations. In that case, a photographic plate on a revolving cylinder had
been considered. Such an observation was performed later by Arnulf
(\cite{arnulf36}).

It is worth noting that a measure of the star flux is actually obtained every time a
column is read out. In standard full-frame CCDs this can be done typically at
frequencies of 10 to 500\,kHz, fast enough for LO work. The fact that these CCDs can
only readout one column at a time makes temporal sampling non-optimal, and
introduces some smearing into LO lightcurve. However, this can be overcome by
compressing the image scale in order to image the star over a few pixels.

In addition, as our technique makes use of a CCD, a flatfield calibration
could be required. Also, residual image motion over several pixels due to
atmospheric turbulence could in principle produce a smearing of the fringes.
However, the compression of the image scale, plus the fact that pixel
sensitivity does not significantly  change along a small CCD portion, minimize
the incidence of this effect.

As SNR plays a key role in data analysis, the choice of a detector with both
high QE and low RON is crucial for data quality. From this point of view, CCDs
look very attractive in terms of QE compared to PEPs.  We see, therefore, how a
full-frame CCD could be used for recording a fast photometry event as LOs. This
detector, far from being a specialized one, is very common among
instrumentation available in all astronomical observatories. Moreover, it turns 
out to be a low-cost satisfactory solution for sub-meter class telescopes of
low-end professional and high-end amateur profile.

\subsection{Signal-to-noise ratio evaluation} \label{snr_evaluation}
Eq.~\ref{sigma_poisson+lognormal+gauss} should be the expression to use when
evaluating SNR for a given detector. However, as the purpose of this paper
is to compare the performance of the technique with other classical detectors,
we do not include scintillation noise the in SNR estimations described
in Sect.~\ref{app}. This will not bias our conclusions, as atmospheric turbulence
affects in the same way both observing approaches.

The SNR for a pure-Poisson detector like PEPs placed in the image plane can be
expressed as:
\begin{equation}
SNR=\frac{N_{\mathrm{*}}}{(N_{\mathrm{*}}+N_{\mathrm{b}})^{1/2}}
\label{snr}
\end{equation}
where $N_{\mathrm{*}}$ and $N_{\mathrm{b}}$ account for number of photon counts
during integration time $\tau$ due to the star and sky background, respectively.
Eq.~\ref{snr} can be reformulated as Sturmann (\cite{laszlo94}):
\begin{equation}
SNR=
\frac{8.9 \kappa^{1/2} F_{\mathrm{*}} D^{2}(D+v\tau)\tau^{1/2}}
{(F_{\mathrm{*}} D^{2}+F_{\mathrm{b}} b^{2})^{1/2}}
\label{snr2}
\end{equation}
where \mbox{$F_\mathrm{*}=1.10\times10^{7}\times10^{-0.4m_\mathrm{V}}$} and
$F_\mathrm{b}=9.95\times10^{6}\times10^{-0.4m_\mathrm{V}^{\mathrm{bg}}}$ are
the extra-atmospheric average photon fluxes for a star of magnitude
$m_{\mathrm{V}}$ and sky background of magnitude $m_{\mathrm{V}}^{\mathrm{bg}}$.
Both correspond to a temperature $T=6000\mathrm{K}$ and all expressed in
[$\mathrm{photons}~\mathrm{m^{-2}}\mathrm{s^{-1}}\mathrm{\AA^{-1}}$] and
[$\mathrm{photons}~\mathrm{s^{-1}}\mathrm{\AA^{-1}}\mathrm{arcsec^{-2}}$],
respectively. The angular extension of the recorded scene projected over the
image plane, measured in [$\mathrm{arcsec^{2}}$], is given by $b$. Finally,
$\kappa$ stands for the product of detector quantum efficiency ($QE$) and a
weighting function $g(\lambda)$ correcting flux for atmospheric extinction and
optical system absorption. A typical value for $g$ at
$\lambda$=6500$\mathrm{\AA}$ is 0.6.

Eq.~\ref{snr2} for the case of CCD turns into:
\begin{equation}
SNR_{\mathrm{CCD}}=
\frac{8.9 \kappa^{1/2} F_{\mathrm{*}} D^{2}(D+v\tau)\tau^{1/2}}
{(F_{\mathrm{*}} D^{2}+F_{\mathrm{b}} b^{2}+\sigma_{\mathrm{CCD}}^2)^{1/2}}
\label{snr3}
\end{equation}
where $\sigma_\mathrm{CCD}$ is expressed in [$\mathrm{photons}$].

For assessing the theoretical SNR performance between PEPs and CCDs, we
consider the following parameters to be input in Eqs.~\ref{snr2} and \ref{snr3}
for either case:
$m_{\mathrm{V}}$$\sim$4, $v$$\sim$0.5$\mathrm{m\,ms^{-1}}$,
$\tau$$\sim$1$\mathrm{\,ms}$, $D$=0.36$\mathrm{\,m}$,
$\sigma_{\mathrm{CCD}}$=5$\mathrm{\,counts}$, $b$=4$\mathrm{\,arcsec}$ and
$m_{\mathrm{V}}^{\mathrm{bg}}$$\sim$10, which is typical during LO events just
beside the Moon. This is close to the observational setting that we present
in Sect.~\ref{obs}. As for the $QE$ of both detector systems, we adopt
from Kristian \& Blouke (\cite{kristian82}) typical values:
$QE_{\mathrm{PEP}}$$\sim$0.15 and $QE_{\mathrm{CCD}}$$\sim$0.70.

The gain $\eta$ obtained by the use of a CCD is:
\begin{equation}
\eta= \frac{SNR_{\mathrm{CCD}}}{SNR_{\mathrm{PEP}}}\sim2.2
\label{gain}
\end{equation}
Eq.~\ref{gain} shows that, in the regime of a moderately bright star, CCDs
with high $QE$ and low $\sigma$ can perform at least as well as PEPs in LO
work. This is less true as we approach fainter stars, where $\sigma$ is dominant
with respect to Poisson noise. However, well before reaching this point the
lightcurve SNR becomes so low that it is in any case impossible to deconvolve for 
instrumental effects ($D$, $\Delta\lambda$ and $\tau$).

It is also worth remarking that in the case of our proposed technique, it
is possible to adjust {\it a posteriori} the size of the integrating aperture. In the
data reduction stage, the user could adapt this size to obtain the optimum SNR 
depending on the actual image motion and on the brightness of the source. In the
case of PEPs, the integrating aperture is fixed and in general this has the
disadvantage, especially for faint sources and in the visual range,  of
introducing a large noise contribution from the background.

Thus, we cannot conclude that CCDs perform in general better LO SNR than PEPs.
However, even with non-optimal temporal sampling inherent to the proposed CCD
column shifting  technique, it is reasonable to consider CCD as potential
detector for LO work in moderately bright star regime.

\section{Observations} \label{obs}

\begin{table*}
\begin{center}
\caption[]{Summary of observed occultation events, as reported in Dunham \& Warren \cite{dunham95}}
\label{ephem}
\begin{tabular}{llllllllllll}
\hline
(1) & (2) & (3) & (4) & (5) & (6) & (7) & (8) & (9) & (10) &
(11) & (12) \\
Source &  Date  & Filter & $\lambda_0\pm\Delta\lambda$ & $\tau$ &
$m_{V}$ & CA & PA & Limb vel. & System & Sep. & PA \\
 &  &  & (nm) & (ms) &  & ($\degr$) & ($\degr$) & ($^{\prime\prime}/s$) & magnitudes & ($^{\prime\prime}$) & (deg) \\
\hline
\object{SAO 77911} & 13-03-00 & R & 641$\pm$58 & 2.344 & 4.6 & 34S & 146 & 0.242 & 5.5 & - & - \\
 &  &  &  &  &  &  &  &  & 6.3 & 0.02$^a$ & 178$^a$ \\
 &  &  &  &  &  &  &  &  & 6.3$^a$ & 1.0$^a$ & 335$^a$ \\
\object{SAO 79031} & 14-03-00 & R & 641$\pm$58 & 2.067 & 4.0 & 63S & 123 & 0.350 & 4.5 & - & - \\
 & & & & & & & & & 4.5$^a$ & 0.10$^a$ & 90$^a$ \\
\hline
\end{tabular}
\begin{flushleft}
$^a$~ Uncertain values: see text in Sect.~\ref{SAO77911}.
\end{flushleft}
\end{center}
\end{table*}

\begin{figure*}
\resizebox{\hsize}{!}{\includegraphics{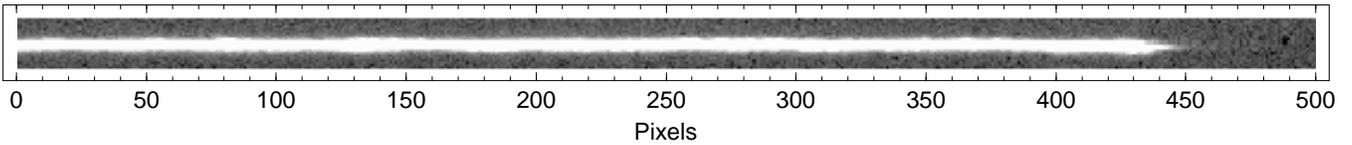}}
\caption{Raw image of SAO 79031 occultation at Fabra Observatory. Strip patch
spans for 1 second of data record, where every 20-pixel column corresponds to
2.067$\mathrm{ms}$}
\label{strip}
\end{figure*}

The current paper examines the small-telescope regime. In particular, a
Celestron 14 inches Schmidt-Cassegrain telescope (hereafter C14) was used to
observe the occultation events reported below. The C14 tube was mounted
parallel to the Mailhat double 38cm-astrograph at Fabra Observatory, Barcelona,
Spain (see Docobo \cite{docobo89} and N\'u\~nez et~al. \cite{jorge92} for a
more specific description of the astrograph). 

Regarding the detector, we employed a Texas Instruments TC-211 CCD, set inside
an SBIG ST8 camera as the tracking chip. This is a full-frame front-illuminated
CCD with 13.75x16 ${\mu}$m pixels and a 192x164 pixel format. Being read out
through a parallel port, its electronic module can operate at 30 kHz with 12
electrons rms RON. With these technical specifications and its high quantum
efficiency (QE peak reaches 70\% at both 650$\mathrm{nm}$ and
730$\mathrm{nm}$), the TC-211 appears to be suitable for fast imaging purposes,
such as tracking and millisecond photometry.

LO data has been acquired by the drift-scanning scheme described in
Sect.~\ref{tdi}. This technique demands an accurate relative timing method
while reading out on a column by column basis (absolute time reference is not a
major concern for stellar diameter determination and binary detection). As the
time sampling required for LO is about 1\,ms, the timing accuracy should be far
below this figure. Accuracies of 1\,$\mu$s are usually achieved either by using
specifically dedicated PC-boards or interrupting CPU internal clock cycle
counter. In our case, the latter option was chosen. Both timing and reading out
procedure was carried out by a DOS-based program called SCAN\footnote{written
by Christoph Flohr, available at
http://home.t-online.de/home/christoph.flohr/tdi\_1e.htm}.

Table~\ref{ephem} summarized the data for two observed  occultation events.
Columns (1) and (2) report source name and observation date. Columns (3) to (5)
correspond to filter name, central wavelength and bandwidth and sampling time.
Visual magnitude is detailed at column (6), while columns (7) to (9) report
position angle (angle along Moon's limb scans the star), contact angle (angle
between lunar motion and scan direction) and angular rate of the event,
respectively. All last three columns are predicted values. Columns (10) to (12)
stands for binary system description: magnitudes of the components, angular
separation and position angle. We have split the components into different
lines.

As an example of strip obtained by the technique described in Sect.~\ref{tdi},
we show in Fig.~\ref{strip} the final part of the \object{SAO 79031}
occultation. In that case, a 20-pixel column is stored every 2.067\,ms on
average. It is important to note here that proposed drift-scanning scheme
allows us to start observation far before predicted LO occultation time. This
gives us more flexibility, since it prevents from eventual errors in such
prediction.

In Sects.~\ref{SAO77911} and \ref{SAO79031} we discuss our results separately
for SAO 77911 and SAO 79031 events.

\section{Data analysis and results} \label{res}

\subsection{Lightcurve model}
The lightcurves were analyzed using a least squares method developed and implemented by Richichi et~al. (\cite{richichi92}). The complete function proposed in such approach is:
\begin{eqnarray}
&&I^{\prime}(t)= F(t)+ \nonumber \\
&&\left[\beta(t)+[1+\xi(t)]\int{\int{I^0(t,\overline{x},\phi) T(\overline{x})S(\phi)d\overline{x}d\phi}}\right]
\label{model}
\end{eqnarray}
where $I^{0}$ is the diffraction pattern of a point-like star, and S and T
describe the instrumental effects discussed in Sect.~\ref{constraints1}. The
background contribution is considered in $\beta$. The signal  can additionally
fluctuate with different temporal frequencies: a low-frequency term is due to
atmospheric turbulence and is represented by $\xi$, and a high-frequency term
is due to residual periodic components in the CCD power supply and cooler
(hereafter pick-up noise), and is accounted by $F(t)$.  

We computed lightcurves from raw strip images by averaging central pixels of
every column and subtracting the background estimated from the outer pixels.
This represents an advantage of the proposed technique with respect to
lightcurves derived from PEP systems, since it incorporates both source and
background level in the recorded signal. In this way, $\beta(t)$ can be discarded
from Eq.~\ref{model}.

\subsection{SAO 77911} \label{SAO77911}

$\chi^{2}$ Ori (HR 2135, BD+20 1233) is a B2 Ia emission line star. Its
angular diameter has been determined by indirect methods three times, with all
values smaller than 1$\mathrm{mas}$ (Pasinetti-Fracassini
et~al.\cite{pasinetti01}). As stated in Table~\ref{ephem}, this is suspected to
be a close binary system, discovered by grazing LO (Reynolds \& Povenmire
\cite{reynolds75}). There is great uncertainty in the separation of the
components derived from that graze, since there were only two rather widely
spaced stations. In addition,  a tertiary component is catalogued. However,
serious doubts about its actual existence have been casted, since if it were
really 1.0$\mathrm{arcsec}$ apart, it would have been resolved by Hipparcos
(Dunham \& Bulder \cite{dunham01b}).

The observation was conducted under partially cloudy conditions, and we are
confident that the SNR of the resulting lightcurve could have been slightly higher
under clear sky. Nevertheless, a visual inspection denotes a clear magnitude drop
at the moment of occultation, and reveals at least the principal diffraction
fringe. The same occultation event was also observed few minutes before by
TIRGO telescope at $\lambda_0$=2.2$\mu\mathrm{m}$.

\begin{figure}
\resizebox{\hsize}{!}{\includegraphics{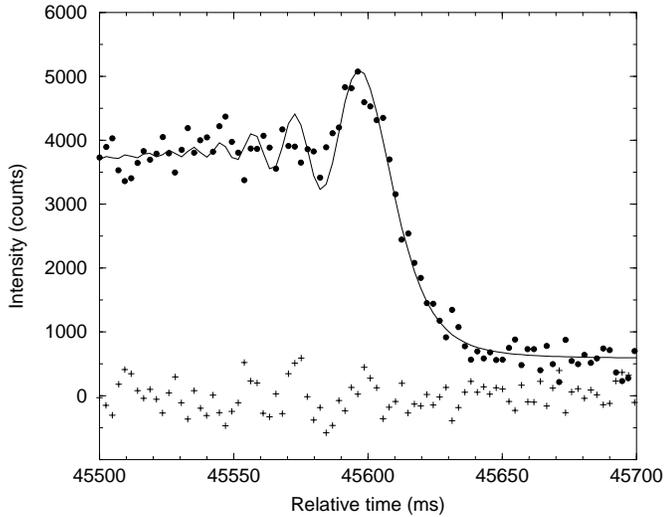}}
\caption{SAO 77911 occultation on March 13th, 2000. Data shown as dots, best
point-like source fit as solid line, residuals as crosses.}
\label{SAO77911_alor}
\end{figure}

We performed a binary fit of both Fabra and TIRGO lightcurves, and no evidence of
binarity was found in any of them. In order to check the good degree of accordance
of SAO 77911 Fabra lightcurve data with single point source model in
Eq.~\ref{model}, we show this in Fig.~\ref{SAO77911_alor} along lightcurve data,
fit and residuals.

\subsection{SAO 79031} \label{SAO79031}

\begin{figure}
\resizebox{\hsize}{!}{\includegraphics{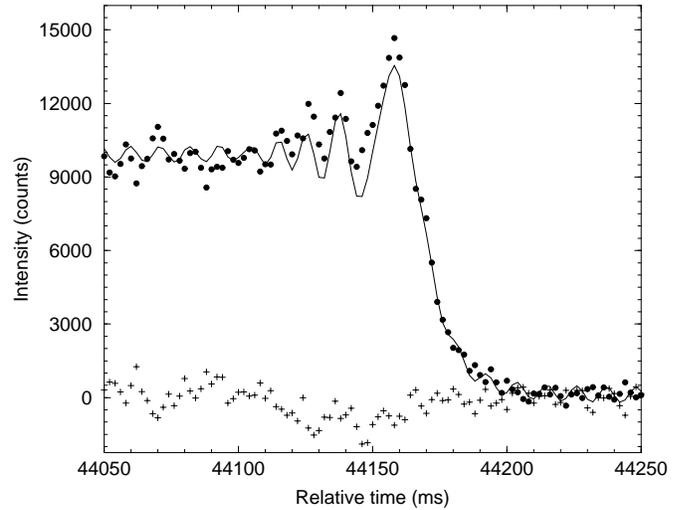}}
\caption{SAO 79031 occultation on March 14th, 2000. Data shown as dots, best
point-like source fit with periodic pick-up noise as solid line, residuals as crosses.}
\label{SAO79031_alor}
\end{figure}

Mekbuda (HR 2650, zet Gem, BD+20 1687) is a Cepheid variable whose
fundamental properties (angular diameter, absolute radius, etc.) have been
studied by several authors (see Table~\ref{check}). It is a multiple system. The
brightest component is catalogued in Dunham \& Warren (\cite{dunham95}) as an
occultation double separated by $0\farcs10$. That was derived from a single
visual observation made with nearly full Moon (Dunham \cite{dunham01a}). On the
other  hand, recent observations performed by modern optical interferometers
clearly discard such duplicity (Nordgren et~al. \cite{nordgren00}). Therefore,
hereafter we will consider \object{SAO 79031} as a single object.

The data are shown in Fig.~\ref{SAO79031_alor}: at least two first diffraction
fringes can be clearly seen. The lightcurve is significantly affected by both
scintillation and pick-up noise. In particular, it has been proven that when
scintillation is not taken into account in the lightcurve model, the derived
angular diameters are biased towards larger values (Knoechel \& von der Heide
\cite{knoechel78}). Our data analysis method, described by Eqn. 13, allowed us to
account for both pick-up noise and scintillation. On the other hand, in our
particular case with small telescope, stellar diameters could be confidently
derived only for very bright and large stars. As shown in Table~\ref{check}, SAO
79031 appears to have an angular diameter smaller than the limiting angular
resolution imposed by our instrumentation ($\phi_m$=4.0$\mathrm{\,mas}$ with R
filter bandwidth). The lightcurve SNR appears to be insufficient to remove such
smearing effect.

Thus, as shown in Fig.~\ref{SAO79031_alor}, we have fitted the lightcurve using
a single point-source model. Superimposed over the theoretical diffraction
curve we have included the 90Hz pick-up noise found in our electrical network.
At the bottom, residuals give idea of the behaviour of scintillation component
of noise. We consider that, taking into account the modest equipment being
used, the fit is in good accordance with the data points.

\begin{table}
\caption[]{Measurements of stellar diameter for SAO 79031}
\label{check}
\begin{tabular}{clcc}
\hline
$\lambda_0$ (nm) & $\phi_{UD}$ (mas) & Observational & Ref. \\
                   &                    & technique     &  \\
\hline
 2200 &  1.6$\pm$0.5  & Lunar occultation & 1 \\
      & 		& (1m telescope)   & \\
 2170 & 1.81$\pm$0.31  & Lunar occultation & 2 \\
      & 		& (4m telescope)   & \\
 2170 & 1.66$\pm$0.16  & Optical interferometry & 3 \\
      & 		& (38m baseline)   & \\
 1670 & 1.88$\pm$0.86  & Lunar occultation & 2 \\
      & 		& (4m telescope)   & \\
 1650 & 1.65$\pm$0.30  & Optical interferometry & 4 \\
      & 		& (104m baseline) & \\
 800 & 1.60$\pm$0.05  & Optical interferometry & 5 \\
     &  	       &  (8-31m baseline)	& \\
 735 & 1.48$\pm$0.08  & Optical interferometry & 6 \\
     &  	       &  (37.5m baseline)	& \\
 450 & 1.66$\pm$0.05  & Optical interferometry & 5 \\
     &  	       &  (8-31m baseline)	& \\
\hline
\end{tabular}
\begin{list}{}{}
\item[1.] Ashok et~al. \cite{ashok94}
\item[2.] Ridgway et~al. \cite{ridgway82}
\item[3.] Kervella et~al. \cite{kervella01}
\item[4.] Lane et~al. \cite{lane00}
\item[5.] Mozurkewich et~al. \cite{mozurkewich91}
\item[6.] Nordgren et~al. \cite{nordgren00}
\end{list}
\end{table}

\section{Summary and final remarks}

A new approach for observing LO has been described. We demonstrate that CCD
drift-scanning at millisecond rate turns to be a viable alternative for LO
observations. In our case, given the small telescope used, the SNR appears to
be insufficient for deconvolving instrumental smearing at the level required
for angular diameter studies, but we estimate that the technique yields enough
spatial information for performing binary detection work.

We remark that the proposed technique can be applied to any CCD which supports
charge shifting at tunable rate. This can be done in nearly all CCDs of
professional profile and in a large number on the amateur market.

In the visible domain, the same technique extended to observations performed
with 1-2\,$\mathrm{m}$ telescopes would increase SNR and, therefore, enable to
deconvolve most of the smearing caused by filter bandwidth and allow stellar
diameter determination.

\begin{acknowledgements}
This work was supported in part by the DGICYT Ministerio de Ciencia y
Tecnolog\'{\i}a (Spain) under grant no. BP97-0903 and AYA2001-3092. O.
Fors is supported by a fellowship from DGESIC Ministerio de Educaci\'{o}n,
Cultura i Deportes (Spain),  ref. AP97~38107939. We would like to express our
gratitude here to Christoph Flohr for making available his program SCAN, which
was crucial for performing observations in drift-scanning mode. We would like to
thank Xavier Otazu for helping in observation tasks. The TIRGO observation of SAO 79031 was obtained by G. Calamai. We thank the referee, Dr. B. Stecklum, for his valuable comments.
\end{acknowledgements}


\end{document}